\documentclass{article}
\usepackage{graphicx} 
\usepackage{svg}
\usepackage[margin=2cm]{geometry} 
\usepackage{amsmath} 
\usepackage{hyperref} 
\usepackage{subcaption}
\usepackage{titlesec} 
\usepackage{float} 
\usepackage{siunitx}
\usepackage{booktabs}
\usepackage{bm}
\usepackage{tikz}
\usetikzlibrary{arrows.meta}
\usetikzlibrary{decorations.markings}
\usepackage{svg}
\newcommand{\DEP}{\text{DEP}}
\newcommand{\EK}{\text{EK}}
\newcommand{\ReA}{Re[\alpha]}
\newcommand{\compleps}{\tilde{\varepsilon}}
\newcommand{\tensor}[1]{\bm{\mathcal{#1}}}

\usepackage{authblk}
\begin{document}
\title{Three-Dimensional Optical and Computational Reconstruction of Colloidal Electrokinetic Flows via Multiplane Imaging and Multiscale Modeling}

\author[1]{Flip de Jong \thanks{These authors contributed equally.}}
\author[2]{Pablo Diez-Silva $^*$}
\author[1]{Jui-Kai Chen}
\author[2]{Ra\'ul P. Pel\'aez}
\author[1]{Sudipta Seth}
\author[3]{Harishankar Balakrishnan}
\author[4]{Bing-Yang Shih}
\author[4]{Senne Fransen}
\author[4]{Wim Van Roy}
\author[4]{Maarten Rosmeulen}
\author[3]{Santi Nonell}
\author[1]{Susana Rocha}
\author[5]{Andrey Klymchenko}
\author[6,7]{Luis Liz-Marz\'an}
\author[3]{Roger Bresol\'{\i}-Obach}
\author[8]{Manuel I. Marqu\'es}
\author[1]{Johan Hofkens\thanks{Corresponding author: johan.hofkens@kuleuven.be}}
\author[2]{Rafael Delgado Buscalioni\thanks{Corresponding author: rafael.delgado@uam.es}}
\author[1]{Boris Louis\thanks{Corresponding author: boris.louis@kuleuven.be}}

\affil[1]{KU Leuven, Department of Chemistry, Molecular Imaging and Photonics, B-3001 Leuven, Belgium}
\affil[2]{Department of Theoretical Condensed Matter Physics, Universidad Autónoma de Madrid, and Condensed Matter Physics Institute (IFIMAC), Campus de Cantoblanco, Madrid 28049, Spain}
\affil[3]{Department of Analytical and Applied Chemistry, AppLightChem, Institut Quimic de Sarria, Universitat Ramon Lull, Via Augusta 390, 08007 Barcelona, Catalunya, Spain}
\affil[4]{imec, Kapeldreef 75, B-3001 Leuven, Belgium}
\affil[5]{Laboratoire de Bioimagerie et Pathologies, UMR 7021 CRS, Université de Strasbourg, 74 Route du Rhin, Illkirch, 67401, France}
\affil[6]{CIC biomaGUNE, Basque Research and Technology Alliance (BRTA), 20014 Donostia-San Sebastián, Spain}
\affil[7]{Centro de Investigación Biomédica en Red, Bioingeniería. Biomateriales y Nanomedicina (CIBER-BBN), 20014 Donostia-San Sebastián, Spain}
\affil[8]{Departamento de F\'{\i}sica de Materiales \& Condensed Matter Physics Center (IFIMAC) \& Nicolás Cabrera Institute, Universidad Autónoma de Madrid, C. Francisco Tomás y Valiente, 7, 28049 Madrid, Spain}

\maketitle

\begin{abstract}
 Sorting, filtering, moving and controlling colloidal particles is crucial in many fields, ranging from chemistry to biology and physics. Dielectrophoresis is an outstanding tool for the manipulation of small particles by AC electric fields, due to its high selectivity and the absence of the need for labels. We use a new theoretical-experimental approach to study the dynamics of fluorescently labeled polystyrene nanoparticles of 200 nm under positive and negative dielectrophoresis conditions. Our multiplane widefield microscopy technique combined with single particle tracking offers real-time ($>$ 100 fps) superresolved visualization of colloidal dynamics in three spatial dimensions. This real-time 3D imaging technique allows the reconstruction of superresolved trajectories, enabling the visualisation of local forces with unprecedented detail. To interpret this data, a dedicated multiscale modeling approach was developed, targeting a direct comparison between theory and experiment. In the current model DEP and electro-osmotic forces were considered. Under positive DEP conditions, this resulted in a very good agreement with experiment. Under negative DEP conditions, the agreement is less clear, indicating the importance of other effects. This illustrates the potential of this combined 3D imaging and modeling approach to validate and refine our theoretical understanding of AC field induced colloidal dynamics. This framework is broadly applicable to other complex fluid or microfluidic motion.
\end{abstract}

\bibliographystyle{unsrt}
\newpage
Selective manipulation of particles is essential in fields such as chemical and biological analysis, food processing, environmental assessment, diagnostics and cell assays \cite{Sajeesh2014}. For example, in biomedicine sorting distinguishes healthy from diseased cells \cite{Fontana2021}, while in materials science precise particle arrangement determines advanced properties, as is the case for photonic crystals or metamaterials \cite{John1987}. This manipulation depends on the particles' physical properties (e.g., size, shape, charge), chemical properties (e.g., material behavior, polarity, solvent/suspension medium), and external force fields (e.g., optical, magnetic, electric)\cite{Zhang2020}. Among the techniques used to manipulate small particles, dielectrophoresis (DEP) stands out for its rapid, label-free, and highly selective abilities \cite{Chen2019}.\\

DEP is the interaction between a polarizable particle and a non-uniform electric field. Particles become polarized under an external field, with an induced dipole moment proportional to the field itself. This leads to a net force whose time average is non-zero, even in alternating (AC) fields, and which can attract or repel particles toward regions of high electric field density. The direction of the DEP force is determined by the net polarizability, which for relatively large particles ($> 100$ nm) is described well by the Clausius–Mossotti factor ($f_\text{CM}$) \cite{Hannay1983}. This factor depends on the dielectric contrast, i.e., the difference between the complex electric permittivities of the particle ($\tilde{\varepsilon}_p$) and the medium ($\tilde{\varepsilon}_m$). If the polarizability of the particle is positive, the particle experiences what is called positive DEP (pDEP) and moves toward the region of higher electric field density (this mode is commonly used to trap particles). Conversely, if the particle’s polarizability is negative, it experiences negative DEP (nDEP) and is repelled toward regions of lower electric field density \cite{Chen2019}. Moreover, the use of electrolyte solutions induces AC electrokinetic flows, which further complicate the dynamics of nanoparticles.\\

Several studies have used DEP to investigate biological elements like DNA \cite{Regtmeier2007}, cells \cite{Farasat2022}, bacteria \cite{Yang2012}, viruses \cite{Lapizco2007,Nakano2016}, and nanostructures like isotropic nanoparticles \cite{Chen2019,Lapizco2007,Gierhart2007}, nanotubes \cite{Duchamp2010}, and nanowires \cite{Kim2010}. Despite its successful application in these areas, accurate characterisation and simulation of the elektrokinetic behavior of small particles in solutions under alternating electric fields remains an open problem. Theoretical and simulation models including not only DEP forces but also the effect of electrokinetic flow can predict particle dynamics in a certain parameter range, usually for large micron size particles, but the problem is severely multiscale in time and space.\cite{Sarno2021}  While techniques such as impedancec\cite{Nakano2016, Suehiro1999, Suehiro2003,  Reale2019}, light scattering/extinctionc\cite{Burt1989, Talary1994, Midelet2020}, and fluorescence correlation spectroscopy \cite{Froude2010} have been used to measure the accumulation or depletion of colloids under electric fields, microscopy offers a more detailed picture, resolving the spatial position of particles and their motion. Reports differ regarding the size of the studied particles, the effect that was investigated and the light microscopy approach that was employed. In this paragraph we only focus on the reported microscopy techniques. Morgan and coworkers were among the first to use fluorescence microscopy to visualize dielectrophoretic trapping of sub-micrometer particles \cite{Hughes1998, Green1999, Ermolina2005, Bakewell2006} and gave an instructive overview of their AC electrokinetics.\cite{green_ac_2000} Fluorescence microscopy visualizing one plane (2D) remains the go-to method to visualize small labeled nanoparticles under electric fields.\cite{Williams2009, Bakewell2013, Rohani2014, Bakewell2015} Additional information can be extracted by recording the position of single particles over the frames of a video, focusing for example on trapping statistics at the electrodes \cite{Bakewell2006, Bakewell2003} or on single particle trajectories for velocity analysis \cite{Su2014, Crowther2019}. Fewer reports use a 3D approach. If imaging speed is not important, a 3D image can be obtained by a simple Z-scan, where planes are imaged sequentially.\cite{Bhatt2005} Also hybrid methods are used, where a movie of the motion in the most important plane is recorded and complemented by a stationary Z-scan.\cite{Shin2023} Real-time 3D measurements can yield the full particle dynamics and if the concentration of imaged particles is sufficiently high, this allows visualisation of the particle response to the electric field for the full imaged volume. Stanke et al. used a mirror to obtain simultaneous access to a top and a side view of the sample, and used manual tracking of a few particles to obtain their velocities.\cite{Stanke2011} Another approach is to use only one plane that is selected to be close to a wall (for example the electrodes) and then use the degree to which the particle is out of focus to determine their distance from the selected plane, converting 2D light microscopy videos to 3D tracking information.\cite{Crocker1996, Dettmer2014}\\

Three-dimensional imaging combined with single particle tracking (SPT) enables the observation of both in-plane and out-of-plane motion, which is critical for studying DEP-induced trapping, levitation, and sorting. Furthermore, detailed particle tracking yields real-time trajectories and velocities, allowing calculating the forces acting on individual particles. Although previous attempts using interdigitated transducer electrodes \cite{Alnaimat2018}, confocal microscopy \cite{Challier2021} or digital holography \cite{Haapalainen2010} have given some initial hints, these methods have not yet delivered the consistent, high spatial resolution tracking required for reconstructing the motion and speed maps. PSF engineering could be a great approach as it allows imaging over large volumes with little to no compromise on the detected photons. \cite{Badieirostami2010,Shechtman2015} However, it suffers from being applicable only to limited concentrations of particles due to the errors induced by overlapping PSF. This is particularly problematic when trying to visualise the capture of a large number of particles on electrodes. Multiplane microscopy (MPM) is a great solution for this problem because it allows imaging of a large volume in real-time without strict constraints on the particle concentration, enabling tracking of a large number of particles and statistical analysis of the localisation and velocities of the tracked particles. \\

Here, we use a custom-designed multi-plane widefield microscopy (MPM) \cite{Geissbuehler2014,Louis2020,Louis2023} to track the 3D motion of nanoparticles in the presence of DEP forces generated by AC electric fields. MPM simultaneously records 8 images (planes) at different z-depths covering a volume of $40\times 60\times 4\, \mu\text{m}^3$, enabling real-time, true 3D imaging. We use this technique to track 200 nm polystyrene (PS) nanoparticles under dielectrophoresis conditions in 3D and map the AC electric field-induced flows around a quadrupolar electrode (a design commonly encountered in literature) \cite{green_ac_2000}. To understand the physical effects dictating nanoparticle behaviour a dedicated modeling approach was developed to qualitatively and quantitatively explain the experimental results. Briefly, we computed the electric field and the dielectrophoretic force using finite differences, and we solved the electrokinetic flows with a novel formalism based on particle discretization of the fluid and Green propagator.\\

We show experimental and modeling results for two conditions, corresponding to pDEP and nDEP condition. Particles are visualized trapped in the space between the electrodes, on top of the electrodes and moving towards or away from the device. This comprehensive picture provides an unprecedented level of detail in the field of electrokinetics and is applicable to any electrode geometry as well as other colloidal systems. This method can contribute to a deeper understanding of electrokinetic and other effects at the nanoscale, which can lead to advances in particle manipulation and sorting across disciplines including physics, chemistry, microelectronics, biology and medicine. 

\section{Theory and modeling}
\begin{figure}[h]
\includegraphics[width=1\linewidth]{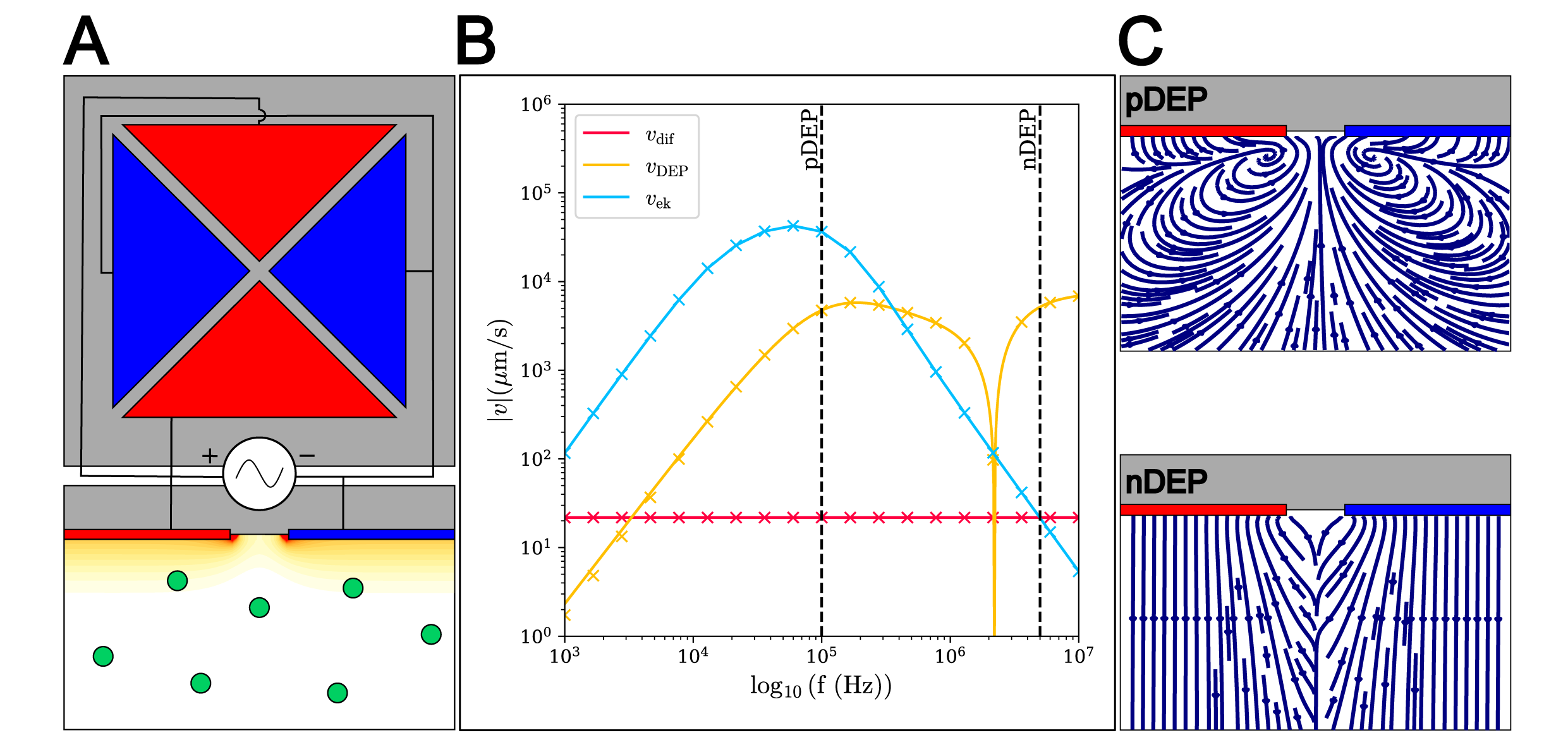}
\caption{(A) Sketch of the quadrupole (top and side view). Red and blue colors represent that each electrode is connected with a phase shift of $\pi$. Side view also includes contour plots of the normalized electric field intensity. Higher yellow color intensity indicates higher electric field intensity, showing that the highest intensity is at the edges of the electrodes. (B) The velocities associated to dielectrophoresis $v_\DEP=\mu F_\DEP$, AC electro-osmosis $v_\EK$ and the characteristic diffusion velocity $v_\text{diff}=\mu kT/a$ with $\mu=(6 \pi \eta a)^{-1}$ the NP mobility and $a$ the NP radius. These values are obtained from theoretical modeling, close to the gap between the electrodes. (C) Schematic representation of streamlines at both working frequencies. The top panel shows the combination of pDEP and electrokinetic flow and the bottom panel the streamlines expected for pure nDEP.}
\label{fig:1}
\end{figure}

The experimental system consists of a microfluidic chip of four metallic plate electrodes over a dielectric (silica) substrate and inside an electrolyte fluid (water). Figure \ref{fig:1} shows the top and side view of a schematic representation of this system. The base of the electrodes is $100 \, \mu\text{m}$ long, the separation between electrodes is $600 \, \text{nm}$, and the height of the electrode above the dielectric is $50 \, \text{nm}$. Since this height is smaller than the NP radius, we simplify the model considering the electrode completely flush with the surface. This approximation may affect the dynamics near the walls, but should have limited effect on the description of the behavior in the bulk. The four electrodes are connected to an AC voltage of amplitude $V_0$ and frequency $\omega$. Adjacent plates present a phase difference in voltage of $\pi$. Inside this system, we introduce spherical polystyrene (PS) dielectric nanoparticles (NPs) of radius $a=100$ nm, and we study their dynamics.\\
The main physical effects that are involved in the nanoparticle motion are the DEP force, that arise from the inhomogeneous electric field created by the electrodes; and the electrokinetic (EK) flow created close to the electrodes due to AC electro-osmosis, i.e. the electrolyte response to the AC field, which is well described in \cite{castellanos_scaling_2003}. The EK flow has a fast oscillatory component (frequency $\omega \in [\text{kHz}-\text{MHz}]$) and a stationary component $\mathbf{v}_0({\bf r})$ which drives the electrokinetic flow.  Here we neglect electrothermal currents and ion currents created around the NPs, based on the expected scaling of these forces compared to DEP and electro-osmosis \cite{castellanos_scaling_2003} (further justification can be found in the Supplementary Information). The aqueous solution has a viscosity $\eta \approx 0.001$ Pa$\cdot$s and density $\rho\approx 10^{3}\text{kg}/\text{m}^{3}$. Due to the small size of the NPs, fluid inertia can be neglected. Moreover, the NP density ($\sim 0.95 \rho$) is similar to that of the fluid and therefore NP inertia and buoyancy are negligible, as it corresponds to the Stokes regime.\\

The time-evolution of the individual NP positions ${\bf R}\equiv \left\{{\bf R}_i\right\}$ can therefore be described by Brownian Dynamics (BD) (we assume that the NP concentration is small enough to neglect hydrodynamic coupling between NPs). The self-mobility of each NP is $\mu = (6\pi\eta a)^{-1}$ and its diffusion coefficient is $D=k_BT \mu$.
\begin{equation}
d\mathbf{R} = \mathbf{u}_0(\mathbf{R}) dt + \mu\mathbf{F} dt + \delta \mathbf{\tilde{R}} + O(dt^2)
\label{eq:brownian}
\end{equation}
The vector of random NP displacements  $\delta \tilde{{\bf R}}$  satisfies the fluctuating dissipation relation $\langle \delta \tilde{{\bf R}} \delta \tilde{{\bf R}}^{T}\rangle= 2D dt$. To that end we use the standard ansatz $\delta\mathbf{\tilde{R}} = \sqrt{2D} \, \delta \mathbf{W}$ built from independent Wiener increments  $\langle \delta \mathbf{W} \delta \mathbf{W}^T\rangle = {\bf I} \,dt$ (with ${\bf I}$ the identity matrix). Consistent with the considered dilute regime, particle-particle (excluded volume or dipolar) interactions are not considered and the forces on NPs ${\bf F}$ correspond to DEP forces and excluded volume interactions with the system's walls. On the other hand, $\mathbf{u}_0(\mathbf{R})$ corresponds with the fluid velocity on a point $\mathbf{R}$. In Brownian dynamics, nanoparticles follow any external flow of the fluid. In our particular case, this corresponds to EK flow.\\

Both DEP forces and the EK flow stem from the AC electric field. To model the AC electric field created inside an electrolyte we have to consider that the free charge density only differs from zero inside the electric double layer (EDL) adjacent to the electrode surface. In the bulk, the fluid is electroneutral and one needs to solve the Laplace equation for the electrostatic potential $\nabla^2 \phi = 0$ (here $\phi({\bf r})$ is a (complex) phasor field representing the time-dependent potential $\mathrm{Re}\left[\phi({\bf r}) \,\exp[i\omega t]\right]$). At the dielectric walls (bottom of the trench between the electrodes) the normal electric field is zero $\nabla \phi \cdot {\bf n}=0$, which means that any current flows through the dielectric. To derive the boundary condition (BC) at the electrode surface we follow Ramos et al. \cite{castellanos_scaling_2003,green_ac_2000} and consider that the EDL width (the Debye length $\lambda_D$) is small compared to any other system's scale $L$. This condition $\lambda_D/L<<1$ is valid for large enough ionic strength and particle size (as discussed later) and permits to treat the charged Debye layer as a capacitor providing a consistent effective BC applied at the thin layer,
\begin{equation}
    \sigma_m \,\nabla \phi \cdot {\bf n} = \frac{1}{Z_{DL}}(\phi - V_0) 
    \label{eq:CC_DL}
\end{equation}
The gradient in Eq. \ref{eq:CC_DL} is taken along the normal-to-wall coordinate ${\bf n}$. The medium's conductivity is  $\sigma_m$ and $Z_{DL}$ is the impedance associated to the EDL which we assume to be a perfect capacitor, i.e. $Z_{DL}^{-1} = i\,\omega C$ with capacitance $C = \varepsilon/\lambda_D$ \footnote{As mentioned by Ramos et al. \cite{green_fluid_2002} it is also possible to use a generalized model with a minor resistance component}. The Debye length $\lambda_D$  for our 1:1 electrolyte is $\lambda_D = \left[\varepsilon k_BT/(n_0 q^2) \right]^{1/2}$ with $q=e$ the electron charge, $T\simeq 300\text{K}$ the temperature, $\varepsilon = \varepsilon_r\varepsilon_0$ the absolute dielectric permittivity of the medium and $n_0$ the ion concentration.  Introducing the dimensionless frequency $\Omega = \omega \varepsilon L/(\sigma \lambda_D)$,  the BC of Eq. \ref{eq:CC_DL} becomes $ \nabla  \phi \cdot {\bf n} = i\Omega(\phi - V_0)$.  Thus, it is evident that  if $\Omega \ll 1$ the electrolyte will fully screen the electric field within the EDL, yielding a Neumann BC  $\nabla \phi \cdot {\bf n}=0$.  In the other limit, $\Omega \gg 1$, the ions do not have time to adapt to the fast electric field and the electrolyte will behave like a vacuum ($\phi = V_0$).  We use a regular mesh to solve the electrostatic potential in the relevant central portion of the experimental $100$ $\mu\text{m}$-side quadrupole. The box sides ($L_x=L_y=20 \mu\text{m}$ and $L_z=5\mu\text{m}$) are far enough to consider Neumann BCs $\nabla \phi \cdot {\bf n}=0$ at $x=\pm  L_x/2$ and $y=\pm L_y/2$ and $z=L_z$. 
\\\\
The resulting electric field ${\bf E}=-\nabla \phi$ leads to the DEP force on the NPs. A particle with polarizability $\alpha$ under an external field ${\bf E}$ presents an induced dipolar moment $\alpha {\bf E}$ and an instantaneous electrostatic force $\alpha {\bf E}\cdot \nabla {\bf E}$. In an inhomogeneous AC field this  yields a non-vanishing temporal average, which in terms of phasor fields is given by
\begin{equation}
    \mathbf{F}_\DEP = \frac{1}{2}\ReA\mathbf{\nabla}|\mathbf{E}^2|,
\end{equation}
which is the DEP force written in conservative form (i.e. as the gradient of an effective potential energy $U_\DEP = -(1/2)\ReA|E^2|$). Notably, the fate of the NP depends on its polarizability: if $\ReA > 0$ the DEP force points towards the maxima of the electric field intensity $E^2$ leading to pDEP.  Conversely, if $\ReA < 0$, the DEP force drives the particle towards low field regions leading to nDEP. For the moderately large size of our NPs ($\sim100$ nm) it is possible to use the Clausius-Mossotti relation to calculate $\alpha(\omega)$ and  include the contribution from the ion conductivity $\sigma$ according to the Maxwell-Wagner approximation, leading to $ \alpha = 4\pi a^3 \varepsilon_m \varepsilon_0 f_{\text{CM}}$ where 
the Clausius-Mossotti (CM) factor $f_{\text{CM}}$ is
\begin{equation}
    f_{\text{CM}} = \frac{\compleps_p - \compleps_m}{\compleps_p + 2\compleps_m}
\end{equation}
with subscripts $p$ and $m$ referring to the particle and the medium, respectively, and the complex permittivity defined as $\compleps = \varepsilon - i\sigma/\omega$. Note that $f_\text{CM}$ determines the sign of the DEP force, resulting in either pDEP or nDEP.\\   

AC electrokinetic flow is nicely explained by  Ramos et al. \cite{green_fluid_2002}. A net slip flow is created inside the Debye layer due to the non-vanishing electric field component in the horizontal directions  ${\bf s}=(x,y)$, parallel to the electrode plane. Within the charged layers adjacent to the electrodes  $\nabla  \phi \cdot \mathbf{s} \neq 0$, pushing the ions and creating a net (averaged) tangential flow with a "slip" velocity 
\begin{equation}
    \left<\mathbf{v}_{\text{slip}}\right> = -\Lambda \frac{\varepsilon}{4\eta} \mathbf{s} \cdot \nabla(|\phi -V_0|^2) \;\mathbf{s} 
    \label{eq:vslip}
\end{equation}
Where $\Lambda$ is the ratio between the diffuse layer potential drop and the total double layer potential drop \cite{green_fluid_2002}. This parameter is commonly used to fit theory to experiment. In this work, a $\Lambda$ value between $0.3 -0.8$ gives a good agreement with experiments. To allow a direct comparison between different conditions, we fixed $\Lambda=0.5$ for all calculations.\\

This tangential flow propagates by continuity to the bulk creating an electrokinetic flow  ${\bf v}_0({\bf r})$ composed of pairs of counter-rotating vortex, which is discussed in Results and Discussion. The standard way to solve this flow consists of imposing the slip BC in Eq.(\ref{eq:vslip}) at the electrodes' surface and solve the Navier-Stokes equations in a finite box, meshed via finite differences to obtain ${\bf v}_0({\bf r})$ everywhere. However, the standard route introduces some technical problems: the system is unbounded in the plane directions (x,y) and a finite box will always introduce hydrodynamic finite size effects. Moreover,  new electrode designs require re-meshing and modifying the BCs. Also note that the slip BC in Eq.(\ref{eq:vslip}) corresponds to a coarse-grained view of the flow within the Debye layers, whose real structure locally resembles a Blasius flow layer with no-slip at the electrode surface and a current above it. Here we use an alternative formalism based on particle discretization of the fluid and Green propagator which fully circumvents these problems and permits a seamless connection with NP dynamics, as schematically represented in Figure \ref{fig:2}C(ii).\\

The slip flow is introduced through a force density distribution $\mathbf{f}(\mathbf{r})$, localized within the Debye layer ($\lambda_D \approx 40\,\text{nm}$), which propagates flow into the bulk via the Blake Green's tensor $\boldsymbol{G}_B(\mathbf{r},\mathbf{r}')$ \cite{swan_simulation_2007,libmobility}. This procedure enforces a realistic no-slip boundary condition (BC) at the electrode surface ($z=0$), yielding a fluid velocity
\begin{equation}
    \mathbf{v}_0(\mathbf{r}) = \int_{V'} \boldsymbol{G}_B(\mathbf{r},\mathbf{r}')\, \mathbf{f}(\mathbf{r}')\, d\mathbf{r}',
\end{equation}
which properly spreads over the unbounded domain.\\

In the discrete particle formulation, the force density is modeled by assigning forces to virtual particles distributed along the Debye layer (orange blobs in Fig.~\ref{fig:2}C). These forces $\mathbf{F}_i$ acting on virtual particles generate a velocity on any other particle $\mathbf{v}_0(\mathbf{R})$ at any position $\mathbf{R}$, mediated by the mobility tensor $\tensor{M}$, so the total velocity is the sum over this hydrodynamic coupling of all the particles of the system
\begin{equation}
    \mathbf{v}_i(\mathbf{R}_i) = \sum_j \tensor{M}(\mathbf{R}_i,\mathbf{R}_j)\,\mathbf{F}_j.
    \label{eq:hydro}
\end{equation}
To impose the slip BC within the Debye layer we have to know the velocity of the virtual particles of the Debye layer, which is calculated by interpolating Eq.(\ref{eq:vslip}) on the virtual particles. The immersed boundary method (IBM) \cite{UAMMD2025,peskin_flow_1972} is used to interpolate the fluid velocity at the particle.\\

Imposing a target velocity $\mathbf{v}^{\text{slip}}_i$ on these particles requires determining the forces that generate it. According to Eq.(\ref{eq:hydro}), this leads to the linear system
\begin{equation}
    \mathbf{v}^{\text{slip}}_i = \sum_j \tensor{M}(\mathbf{R}_i,\mathbf{R}_j)\,\mathbf{F}_j,
    \label{eq:fminv}
\end{equation}
whose solution provides the force distribution responsible for the slip flow.\\

We solve Eq.~(\ref{eq:fminv}) using the GPU-accelerated library \texttt{Libmobility} \cite{libmobility}, which computes the mobility tensor $\tensor{M}$ for different geometries. Instead of explicitly inverting $\tensor{M}$ (prohibitively expensive for large systems), we employ a GMRES iterative solver \cite{saad1986gmres}, which only requires evaluating products of the form $\tensor{M}\mathbf{v}$, efficiently provided by \texttt{Libmobility}. We use the Blake tensor in open space \cite{swan_simulation_2007} to ensure the no-slip BC at the electrodes, and the slip velocity is imposed at the midpoint of the EDL (20\,nm above the wall).\\

Once the force distribution is obtained, the resulting bulk flow is sampled by placing additional virtual particles (gray blobs in Fig.~\ref{fig:2}C) and spreading their velocities onto a 3D regular mesh using the IBM. This unified procedure is used consistently for both electrokinetic (EK) and dielectrophoretic (DEP) simulations. A detailed description of the method is provided in SI~XX.

\section{Results and discussion}
\subsection{Modeling results}
The outcome of the theoretical analysis is summarized in Fig. \ref{fig:1}.
Figure \ref{fig:1}B compares the velocities associated with the three main transport mechanisms: dielectrophoresis  ${\bf v}_\DEP({\bf r})=\mu F_\DEP({\bf r})$, EK flow ${\bf v}_\EK({\bf r})$ (both evaluated at a distance z=100nm from the electrode surface) and diffusion. To compare the first two with the thermal noise, we define the diffusion velocity $v_\text{dif}=\mu k_BT/a$ by calculating the time it would take for a particle to diffuse a mean square displacement equal to the squared radius of the particle $a^2$. We used the relative permittivity of the solvent (water) $\varepsilon_m \approx 78.5$ and NP (polystyrene) $\varepsilon_p \approx 2.5$. For these values, one obtains a purely dielectric response  at high frequencies, yielding nDEP with $f_{\text{CM}} \approx -0.5$. By contrast, conductivity transport, determined by $\sigma$, dominates at low frequencies. Therefore, we expect an intermediate cross-over frequency $\omega_{cr}$ at which $f_\text{CM}(\omega_{cr})=0$. A condition for obtaining pDEP at low frequencies if $\varepsilon_m > \varepsilon_p$ is that $\sigma_p > \sigma_m$, however, the conductivity of water at 0.1 mM KCl ($\sigma_m = 0.0015$ S/m) is much higher than the electrical conductivity of polystyrene ($\sim 10^{-20}$ S/m), which would lead to nDEP. It should however be considered that the additional contribution from ion conduction across the EDL shell surrounding the NP significantly increases $\sigma_p$. Experimental measurements of carboxylate-modified particles immersed in a buffered solution of $0.1m$M KCl pH 6  indicate that $\omega_{cr} \in [2.3 - 2.4] \text{MHz}$. This corresponds to  $\sigma_p \approx 0.013$ S/m, which is in agreement with the theoretical calculations made by H. Zhao et al. \cite{zhao_polarization_2009} for a particle of 108 nm with bulk conductivity  $\sigma_m = 0.0015$. Therefore we use $\sigma_p = 0.013$ S/m in our calculations. 
\\\\
The vertical dashed lines in Fig. \ref{fig:1}b indicate the frequencies used in our experiments, showing pDEP at 100 kHz and nDEP at 5 MHz. Note that EK flow is dominant at low frequencies and negligible at high frequencies. In particular, at high frequencies the EK velocity becomes smaller than the “Brownian velocity” $v_\text{diff} = \mu kT/a \approx 20,\mu\text{m/s}$ (although it is important to recall that the average Brownian velocity is zero, whereas EK flow represents a systematic drift). When ${\bf v}_\text{EK} \approx 0$, the steady-state NP density is close to the equilibrium distribution associated with the DEP potential, 
\begin{equation}
    P\text{st} \approx P_\text{eq} \propto \exp(\beta U_\text{DEP})
    \label{eq:prob}
\end{equation}
Figure \ref{fig:1}c shows theoretical trajectories of the studied NPs at both operating frequencies. At 100 kHz, the particles experience pDEP, and the EK flow is very intense. DEP pulls particles toward the electrode edges, while the EK flow generates two counter-rotating vortices along the dielectric corridor. The total motion results from the combination of both effects. In contrast, at 5 MHz NPs experience nDEP and the EK flow is much weaker (below the level of thermal noise). DEP pushes the NPs away from the electrodes to the region where thermal noise dominates, and NPs are expected to move randomly.\\
\begin{figure}[H]
    \centering
    \includegraphics[width=1\linewidth]{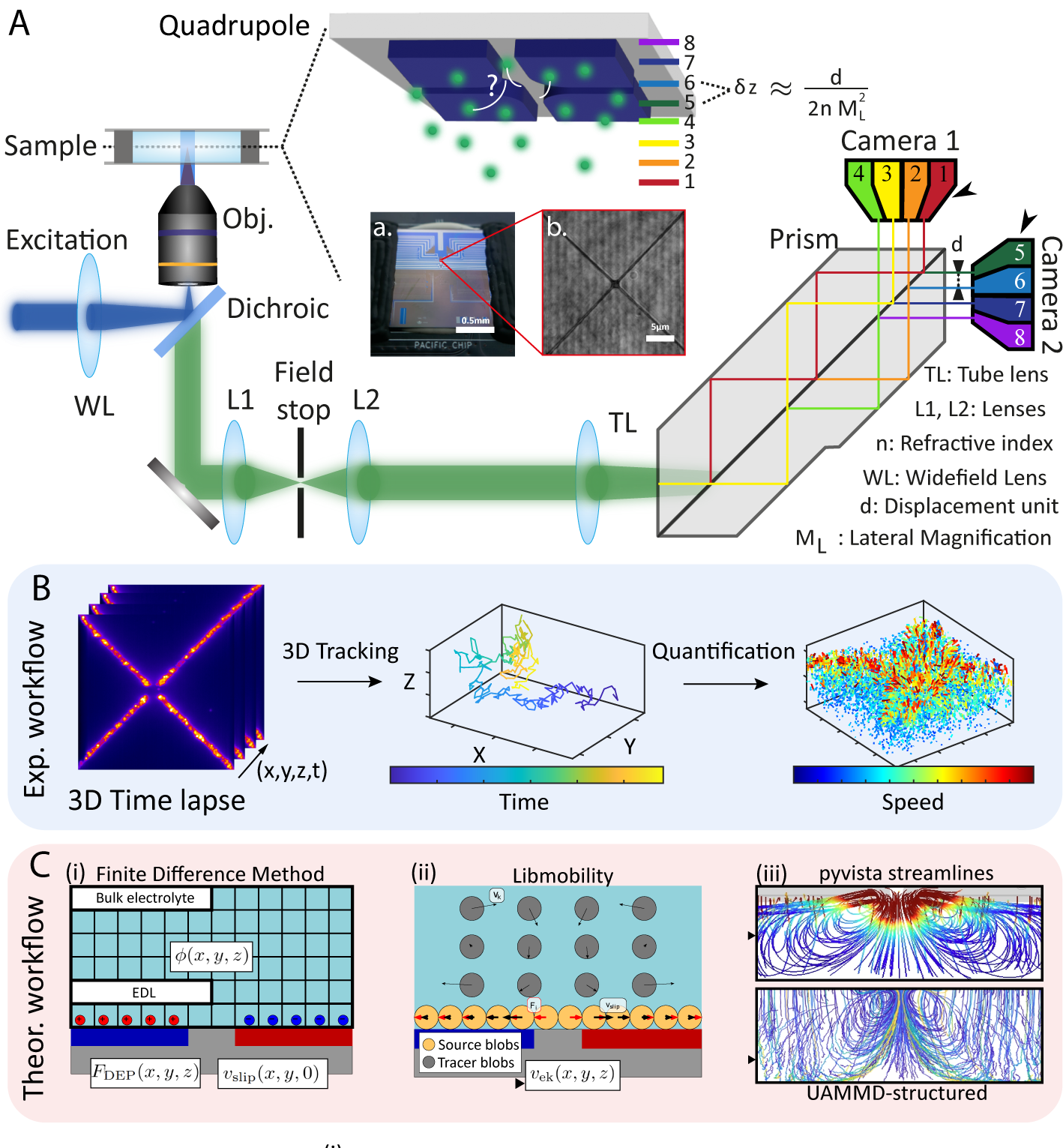}
    \caption{3D imaging experimental setup with experimental and theoretical workflows: A) a 488 nm laser is focused on the back aperture of the objective to generate widefield illumination. The light emitted from the sample is collected via an 8f system. A prism beamsplitter splits the image in 8, each of the split paths having a different optical path length, which leads to eight images focused at different depth inside the sample, resulting in a true volumetric image. The images are collected 4 by 4 by two synchronized sCMOS. a) Photograph of the microfluidic device containing the quadrupole electrode. b) scattering image of the quadrupole electrode under the microscope. B) Experimental workflow: A 3D time lapse of the imaged volume si acquired under application of an alternating electric field. Individual particles are tracked in 3D, and their speed is calculated and mapped out in all dimensions. C) Theoretical workflow: The space is discretized and the finite difference method is used to calculate electric potential and $F_\DEP$ in all the system. The $v_\text{slip}$ is propagated to the bulk to calculate EK flow in all the system using libmobility \cite{libmobility}. Then the total speed field is calculated with pyvista streamlines \cite{pyvista} and the NPs dynamics are solved with UAMMD-structured Brownian solver \cite{UAMMD_structured}}.
    \label{fig:2}
\end{figure}
 We will now present the multiplane microscopy setup for 3D visualization and then compare the experimental findings with the theoretical predictions. In the case of pDEP we find an excellent agreement, while some discrepancies are observed for the case of nDEP. The physical origin of such discrepancies can be related to effects that are not included in the present model, for example higher order effects, particle-particle interactions or electrothermal flows around the NPs. 

\subsection{Multiplane microscopy and dielectrophoresis setup}
To experimentally  investigate  NP motion under DEP, we employed a simple quadrupolar electrode (total area $(100\times 100)\, \mu\text{m}^2$,  with a separation between the electrodes of 600 nm and a depth of the trench (inter-electrode corridor) of 50 nm) connected to a PCB board and driven by a Keysight waveform generator (EDU33212A) which was controlled by a homemade LabVIEW program. Details of the setup are provided in SI (SI Note XX) and the quadrupole device is schematically depicted in Figure \ref{fig:2} along with a picture of the chip (\ref{fig:2}A) and a backscattering image of the quadrupole (\ref{fig:2}B).\\

For three-dimensional NP tracking, we used our custom-built multiplane microscope \cite{Louis2020}, as shown in Figure \ref{fig:2}A. Briefly, a prism beam splitter divides the detected emission into eight separate images, 4 collected on each of the two synchronized sCMOS cameras. The prism’s geometry creates different optical path lengths between the tube lens and the cameras, conjugating eight distinct z-planes within the sample to 8 different images on the detector. The focal plane is positioned to have the top plane near the electrode interface, enabling imaging of a volume of $50\times 50 \times 4 \mu\text{m}^3$ at a high frame rate (125 fps). Calibration was performed as previously described \cite{Louis2020} and determined an inter-plane distance of approximately 580 nm (see SI Figure SX), with tracking precision 10 nm (x-y) and 20 nm (z) and accuracy of 20 nm (x–y) and 40 nm (z).\cite{Louis2020} \\

\emph{3D visualization and modeling of 200 nm PS particles under positive and negative dielectrophoresis.} To visualize the dynamics of 200 nm PS particles under DEP, we first established the voltage and frequency regimes for pDEP and nDEP. We then optimized the voltage to ensure that particles did not move too fast nor too slow, allowing accurate visualization at 125 fps. Two conditions were selected: 100 kHz at 0.5 V to show pDEP, and 5 MHz at 1 V where the PS particles experience nDEP. The experimental workflow is schematically described in Figure \ref{fig:2}B. Briefly, a 10 second (1250 frames at 125fps) 3D time lapse was acquired during which an alternating electric field was applied at the desired voltage and frequency. For each movie, we then tracked the particles in 3D, using our custom software (freely available at https://github.com/BorisLouis/3D-image-processing) \cite{Louis2020}. Finally, we determined the speed and combined the data of 10 different movies to increase the statistics and allow mapping of the speed of the particles in the 3D space. 

Using the fast 3D imaging afforded by MPM, we could directly observe how particles move under different conditions around a quadrupolar electrode. Combined with the modeling approach put forward in the previous section, this experimental information can provide a quantitative measure on the DEP force field and accompanying electrokinetic flows, a solid base to validate models. One of the advantages of particle tracking is that it allows one to measure the “instantaneous” particle velocity (more precisely ${\bf v}_p= \Delta {\bf R}/\Delta t$ with $\Delta t= (1/125)\text{s}$ as the sampling frequency of experimental frame). As our dynamic simulations also allow us to calculate the velocity of particles in the fluid, we can quantitatively compare experiment and theory. 

\subsection{Positive DEP}
\begin{figure}[H]
    \centering
    \includegraphics[width=1\linewidth]{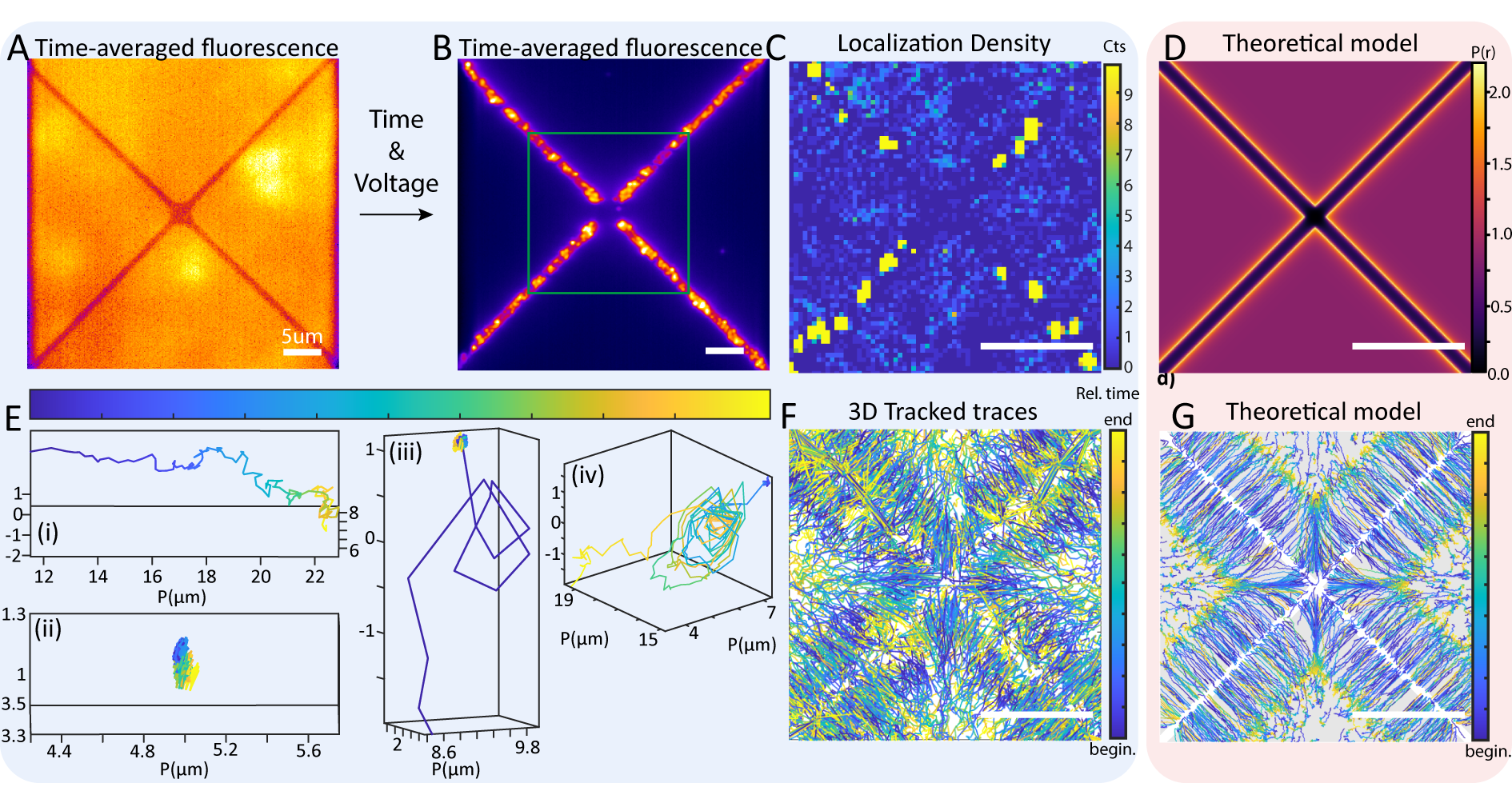}
    \caption{Experimental and theoretical analysis of pDEP. A) Time-averaged image in absence of any voltage showing a homogeneous intensity distribution. B) Time-average image under 100kHz and 0.75V AC field showing preferential location of particles inside the electrode gap where the electric field gradient is strongest. C) Localization density at the electrode combining data from multiple periods of 10 second exposure to 100 kHz at 0.5 V (pDEP). D) Simulation showing the  logarithm of the calculated particle distribution probability under these conditions in arbitrary units E) Representative 3D traces of different typical behavior: (i) ejection from center at 45 degrees w.r.t. the electrode arm, (ii) trapping at the electrode, (iii) moving from bulk solution to electrode gap, and (iv) escape from the trap, moving perpendicular to the electrode arm entering a vortex flow. Time color-code is given relatively for clarity (for each particle, the beginning of its trace is depicted in blue, while later times are depicted in yellow). F) Top-view of all the trajectories showing an overview of the characteristic behavior, notably ejection from the quadrupole center at 45 degrees from the electrode arms and ejection perpendicular to the electrode arm. G)  Top-view of all the traces from the theoretical dynamical simulation showing the predicted behavior in a good match with experimental results.}
    \label{fig:3}
\end{figure}

\emph{Localization density, 3D tracking and theoretical view}
We start by analyzing the experimental results for pDEP at $100 \text{kHz}$ and  $0.5V$. Figure \ref{fig:3}  details the results of  pDEP from both the experimental and the theoretical point of view. Figure \ref{fig:3}A shows the average intensity distribution of a 10 s movie without applied voltage. The intensity distribution is more or less uniform due to the random Brownian motion of the particles. Figure  \ref{fig:3}B shows the same period after multiple cycles applying pDEP frequency, demonstrating a clear accumulation of particles in the electrode gaps. This is expected because the region near the electrode is where the electric field intensity (and consequently $U_\text{DEP}$) is highest.\\

To compare with theory, we looked at the localization density (histogram) of the electrode plane (\ref{fig:3}C) which clearly shows that particles are predominantly attracted to and confined within the gap between electrodes (bright yellow spots), close to the electrode edges. Similarly, the predicted probability density Eq.(\ref{eq:prob}), shown in (\ref{fig:3}D), has a strong maximum at the edges of the electrodes, suggesting an accumulation of particles (yellow strips in Fig. \ref{fig:3}B). A slight difference between the theory and the experiment is that in the experiment the particles seem to be more evenly distributed in the electrode gaps while in the theory the potential is mostly high at the edges. This difference is due to the combination of a few effects. First, in experiments the electrode has a height (50 nm), providing a further geometrical constraint not present in theory. It is also important to realise that description of the potential and moving particles are two different things, which can lead to discrepancies (e.g. a particle moving fast and getting stuck is experimentally plausible, but not accounted for from a pure potential view). We note that trapping of smaller particles (63nm) shows nicely that the accumulation is really at the electrode edge, rather than distributed over the corridor (Figure SX).\\

To fully capture the details of NP dynamics we looked at individual trajectories in 3D. Figure \ref{fig:3}E presents representative traces of particle trajectories during pDEP. These traces are color-coded for relative time, with blue representing the earlier part of the trace and yellow the later part. Figure \ref{fig:3} (i) presents a trace that is ejected away from the quadruple center at 45 degree from electrode arms. Figure \ref{fig:3}E(ii) shows a trace of a particle trapped in the electrode gap. Figure \ref{fig:3}E(iii) shows a particle going from the bulk solution to the electrode, with a loop motion on the way. Figure \ref{fig:3}E(iv) particle was initially trapped, then released, likely due to local flow, it was then stuck in a vortex loop before escaping. This vortex motion and ejection is perpendicular to the electrode arm.\\

Figure \ref{fig:3}F shows a top view of all the particles tracked on all the 10 movies using the same relative time color-coding. Similarly to the individual traces extracted before, we observe several areas exhibiting different particle dynamics. Inside the corridors, a high concentration of yellow traces confirms particle capture in the region of high electric field strength, close to the electrodes. Secondly, perpendicular to the corridors of the electrode cross, particles appear to be ejected, as traces transition from blue near the electrode to yellow further away (corresponding to observation at shorter time close to electrode and at longer time away from electrode). This was found later to be associated to vortex-like flows (see details in next section). Finally, we observed that particles are expelled from the electrode center in four directions at 45° angles with respect to the cross formed by the corridors. This effect was found to be stronger at higher voltage, see SI XX.\\

These experimentally observed dynamics agree well with the theoretically predicted behavior. Figure \ref{fig:3}G shows the simulated trajectories of 2000 NPs under DEP and EK flow conditions. The color code is the same as in the experimental figure (3F), to facilitate comparison. We solved the dynamics using the multiphysics software UAMMD-structured \cite{UAMMD_structured}, which includes a Brownian dynamics solver and allows the user to include an external tabulated force field in the simulation. This force field incorporates the contributions of both effects included in our model. The results show that particles approach the electrode's gap due to DEP, and some of them are ejected by EK flows perpendicular to the electrode arms, resulting in a picture that is fully consistent with the experimental observations.\\

\begin{figure}[H]
    \centering
    \includegraphics[width=1\linewidth]{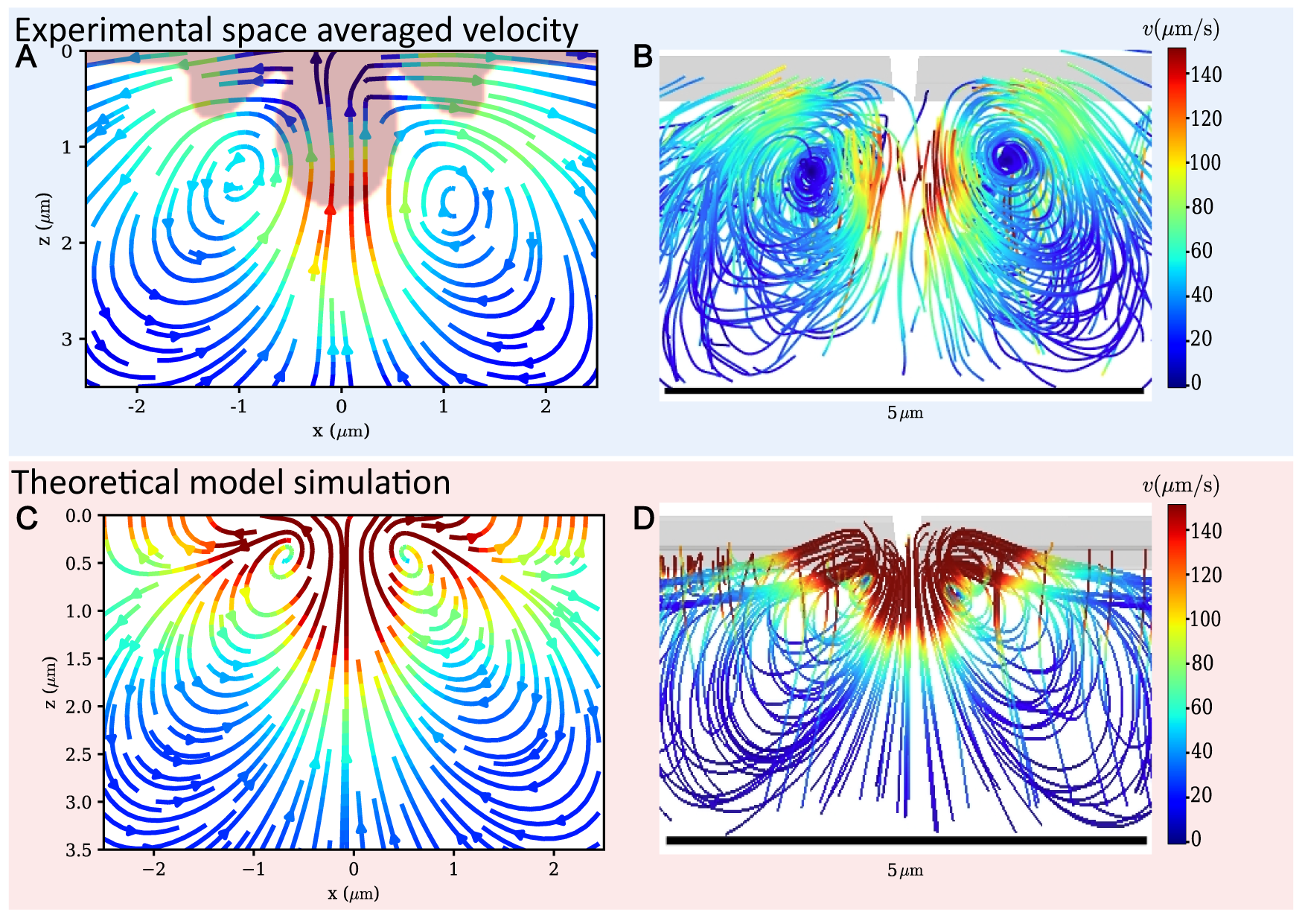}
    \caption{Comparison of experimental and theoretical velocity fields averaged over the corridor direction. The red transparent zone overlaid with the experimental figure indicates where theoretical predicted velocity is higher than $110$ $\mu\text{m/s}$ and proximity to the surface, resulting in inaccurate experimentally recovered velocities.}
    \label{fig:4}
\end{figure}

\emph{3D flow visualization and speed profiles}
Building on the previous section, we use the tracked traces to map the speed of particles in the three dimensional space and compare these velocities with theoretical calculations. Figure \ref{fig:4} shows the particle's motion in a cross-section perpendicular to the electrode gap (see SI X). The aforementioned ejection perpendicular to the electrode in figure \ref{fig:3}F reveals a much more complex motion, like a vortex, when considering all three dimensions. Given the axial direction of the vortex, this could not have been observed easily with a standard microscope. From theory, we understand that this orthogonal flow is essentially driven by AC electro-osmosis, as the DEP force decreases fast- the further the NP move away from the electrodes' surface. This EK flow forms two counter-rotating vortices (Fig. \ref{fig:1}a) unfolding in the normal planes of the inter-electrode corridors and expelling particles to the bulk.

Figure \ref{fig:4} compares the experimental 3D reconstruction of these trajectories, which clearly follow vortex streamlines consistent with the theory. Indeed, an excellent agreement is found both in the vortex dimensions (each one about 2.5 microns in size) and on the typical velocities attained (Fig \ref{fig:4}c). Theory predicts the largest velocities to be at a layer of less than 1 micron adjacent to the electrodes. These velocities are however too high to allow experimental observation in the current conditions. Away from this layer (i.e. about 1 micron away from the electrode surface) experiment and theory agree well on velocities of $\sim 100\mu\text{m}$, which decrease to $\sim 30\mu\text{m/s}$ a further 4 microns away \footnote{Experimental speeds are comprised in the range of $[0-250] \mu\text{m/s}$ s with 95th percentile at $63\mu\text{m/s}$ and a 99th percentile at $145\mu\text{m/s}$.}\\

Close to the electrode center the interaction between EK vortices coming from different electrode arms results in currents tilted 45 degrees from the corridor between the electrodes. This leads to an increase in tracked NPs in these orthogonal directions, which is not immediately apparent from the trajectories using 0.5V in \ref{fig:3}c, but becomes very clear if the amplitude is increased to 0.75V (Fig. in the SI).\\

Fig.\ref{fig:4} shows streamlines of the experimental and simulated velocity fields. As explained higher, the theoretical model for obtaining the velocity field is based on a regular 3D mesh in which each cell is assigned a velocity value. To obtain comparable experimental data, we measured the velocities of all detected nanoparticles at various times by calculating their displacements between two consecutive time frames. We then averaged the velocities within regular boxes of 500 nm per side. In this way, we obtain an experimental velocity field to compare with the theoretical model. Figures \ref{fig:4}A and \ref{fig:4}C show 2D representations of these fields, obtained by averaging over the corridor direction.  These 2D streamlines are calculated using the basic streamplot function in matplotlib in Python. Figures \ref{fig:4}B and \ref{fig:4}D show 3D representations obtained with the PyVista \cite{pyvista} software.
3D renderings of these figures are available at \url{https://padisi.github.io/quadrupole-paper/}.\\

The experimental and theoretical data of particle motion and electrokinetic flow are matching very well. Minor mismatches can reasonably be attributed to experimental limits, as indicated in Fig.\ref{fig:4} In the zone indicated in red, due to the high velocities and the proximity of the geometrical barrier of the surface, no accurate velocities could be recovered.  The combined results from experiment and model qualitatively and quantitatively capture the complexity of the colloidal dynamics in all three dimensions.

\subsection{Negative DEP}
\begin{figure}[H]
    \centering
    \includegraphics[width=1\linewidth]{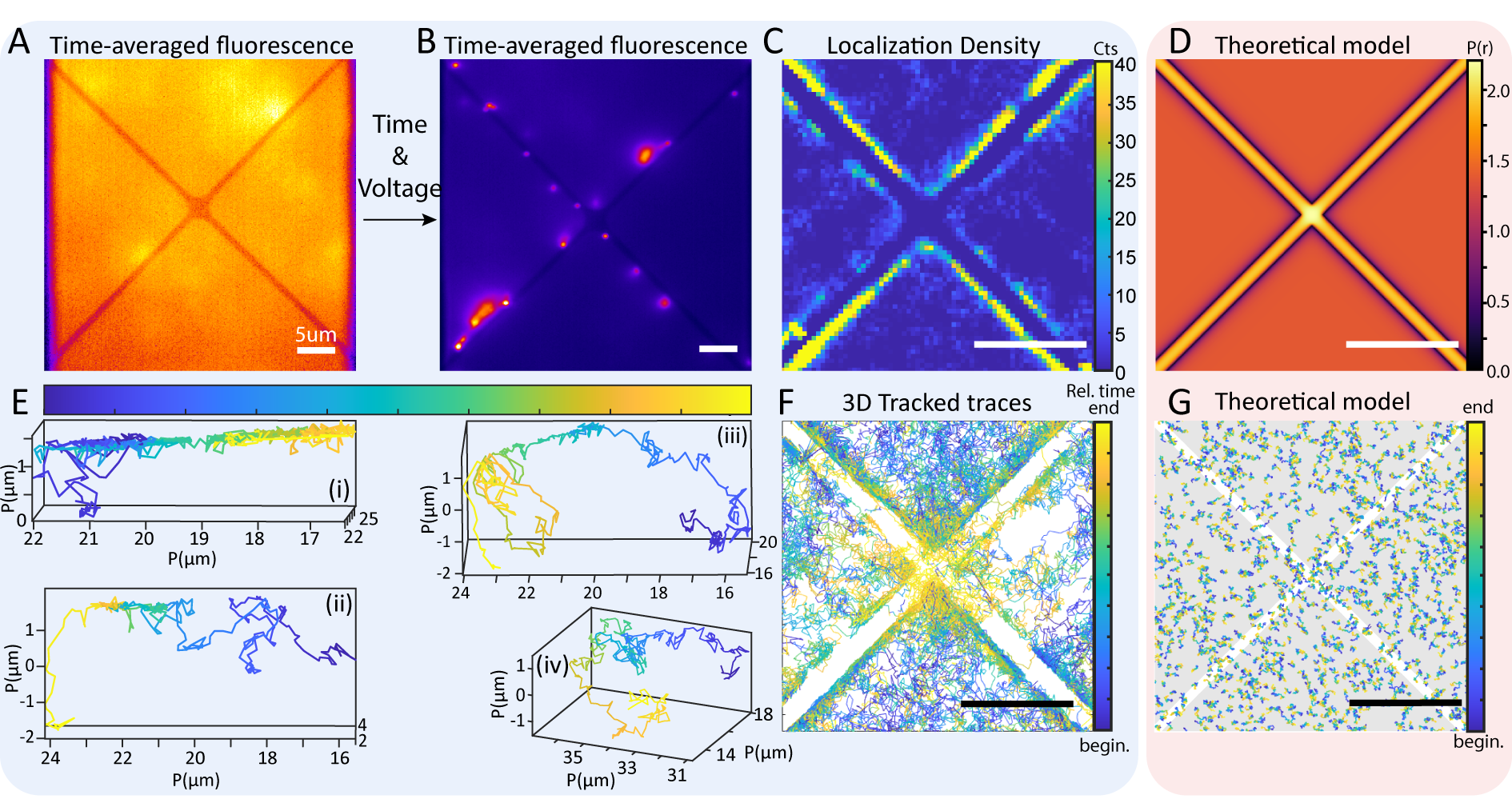}
    \caption{Experimental and theoretical analysis of nDEP. A) Time-averaged image in absence of an AC field, showing a homogeneous intensity distribution. B) Time-averaged image under a 1MHz and 1V AC field, showing preferential location of particles just outside the electrode corridor. C) Localization density combining data from multiple cycles of 10 second exposure to 5MHz at 1V (nDEP). D) Simulation showing logarithm of the particle distribution probability in nDEP in arbitrary units. E) Representative 3D traces of different typical behavior: (i) movement along electrode corridor, moving above the electrode edge, (ii) full cycle, moving from bulk solution to horizontal movement along the electrode gap, and finally repulsion in a vortex toward the bulk after reaching the center of the quadrupole, (iii) shows similar behaviour as (ii) but is expelled more vertically, and (iv) shows Brownian motion in the bulk. Time color-code is given relatively for clarity (for each particle, the beginning of its trace is depicted in blue, while later times are depicted in yellow). F) Top-view of all the detected traces gives an overall view of the the characteristic behaviors, notably the absence of particles inside the gap between the electrodes and the attraction towards the center of the quadrupole G) Top-view of all the traces from the theoretical dynamical simulation showing the predicted behavior, which mostly corresponds to particle being pushed away from the elctrodes and subsequent random, diffusion dominated motion.}
    \label{fig:5}
\end{figure}

Now we turn our attention to the dynamics under typical nDEP conditions, Figure \ref{fig:5} shows results obtained for 5 MHz and 1V. Compared to pDEP, the experimental observation of nDEP dynamics is much more challenging because of its nature: as it tends to repel particles from the electrode to a regime where nDEP is no longer the dominant force, tracking particles under nDEP conditions is difficult. Furthermore, since electro-osmotic flows at this frequency are weaker, other effects (not included in the model) become increasingly relevant.\cite{green_ac_2000}\\

Figure \ref{fig:5}A shows the average intensity distribution of a 10 s movie without applied voltage. The intensity distribution was again observed to be homogeneous, corresponding to Brownian motion of the fluorescent NPs. Figure \ref{fig:5}B shows the same period after multiple cycles of applying the selected nDEP frequency (1V, 5MHz), showing no clear accumulation of particles at the electrode gaps. Instead, we observe an unexpected accumulation of particles just besides the electrode edge, moving above the electrode surface.\\

This view is confirmed by the localization density shown in Figure \ref{fig:5}C depicting a total absence of particles between the electrodes and the presence of particles along the electrode with no clear position (no cluster). This indicate that the particles there are not trapped, but remain in motion. These features can be corroborated by the theoretical predictions of Fig. \ref{fig:5}D, showing the Boltzmann contribution to the NP distribution $P_{eq}\propto \exp[-\beta U_\DEP]$ and the DEP+EK velocity field $[\tilde{{\bf v}}({\bf s},z)\cdot {\bf s}] \,{\bf s}$ projected in the electrode plane at a distance $z=1\mu\text{m}$ below the surface. According to this theoretical evaluation, the DEP potential presents a depletion zone just at the electrode surface, while the gap between the electrodes presents a slightly larger predicted NP density. It should be noted that theoretical calculations have not considered the 3D structure of the 50 nm trench. In the trench interior the electric field is large, so the nDEP particles are strongly expelled from there. Moreover, even in absence of a trench, the 3D density suggests a higher density away from the electrode compared with that in the electrode gap. Therefore the probability of finding particles in the electrode gap is significantly lower than it appears to be in this two-dimensional view. We should also note here that the local concentration of particles detected in the imaged volume is much lower than for pDEP conditions, even though the same bulk concentration was used, which is a clear indication of the repulsion of particles.

When looking in more detail at the observed trajectories of individual representative particles (figure \ref{fig:5}E, we see a several different patterns. Representative traces showcase the four main observed behaviors: (i) Particles are sliding along the electrode edges, around 500 nm below it. (ii) Particles go from the bulk solution and then show sliding along the corridor similar to (i). In this case the particles moving along the corridor often end up at the center of the quadrupole where they are repelled toward the bulk, in this case showing a looping motion. (iii) Particles going from the bulk to the electrode corridor and repelled at the center of the quadrupole downward toward the bulk. And (iv) Brownian motion, as we observed that many particles do not seem to have a significant force acting on them.\\

This also appears clearly when looking at overview figure \ref{fig:5}F where most of the traces seem to show Brownian motion in random directions. It is very noticeable that the gap between electrode seems void of particles, especially considering it is a bottom view, this means that even 3-4$\mu\text{m}$ below the electrode, particles do not easily cross the region of the electrode gap, acting almost like a wall. We also see a concentration of yellow traces (late times) at the center of the quadrupole, indicating that several particles end their trajectory there. However, they do not get trapped at the surface, but are rather dynamically moving above the center after being pushed downward to the bulk, as shown in the individual examples of \ref{fig:5}E.

A main discrepancy between experiment and theory is that the theoretical approach does not capture directly the observed trajectories of nDEP particles that move along the direction of the corridor, 500 nm below the electrodes. However, we will show that the model can capture the horizontal drift towards the quadrupole center. To understand the observed stationary NP distribution one needs to look at the theoretically predicted horizontal NP flows, as the DEP forces act essentially exclusively in the vertical direction while horizontal flows are small and dictated by a delicate balance between advection and diffusion. The streamlines of the DEP+EK flow $\tilde{{\bf v}}({\bf r})$ in the inset of Fig. \ref{fig:1}C corresponding to nDEP indicate a stagnant region at a distance slightly larger than one micron below the inter-electrode corridor. At this region, the downward DEP velocity is balanced by the upward EK flow. Around this stagnant region the field $\tilde{\bf v}$ is mostly horizontal and gently drives the NPs to both sides of the electrode corridors.\\

When analyzing the experimentally observed NP flow in a wider scale we observe the formation of a large vortical structure, which strongly expels (away from the electrodes) any particles located in a region about 1 micron below the center line of the quadrupole  (see Video * in Supp. Info.).  By continuity, the stationary current requires a counter-flow balance which brings particles towards the quadrupole center. Part of this current is due to NPs moving towards the quadrupole center along the diffusive layers adjacent to the electrode arms. As a consequence the NP density is larger just below the quadrupole center, as shown in Fig. \ref{fig:5}D (theory) and also revealed from the (end-trace) yellow region around the center in Fig. \ref{fig:5}C (experiment).\\

The low strength of the EK flow (Fig. \ref{fig:1}b), with velocities in few $\mu\text{m/s}$ range is comparable to the characteristic diffusion velocity $v_\text{diff}=D/a$ (with $D=\mu kT\approx 2.2 \mu\text{m}^2/\text{s}$ the NP diffusion coefficient).\\
\begin{figure}[H]
    \centering
    \includegraphics[width=1\linewidth]{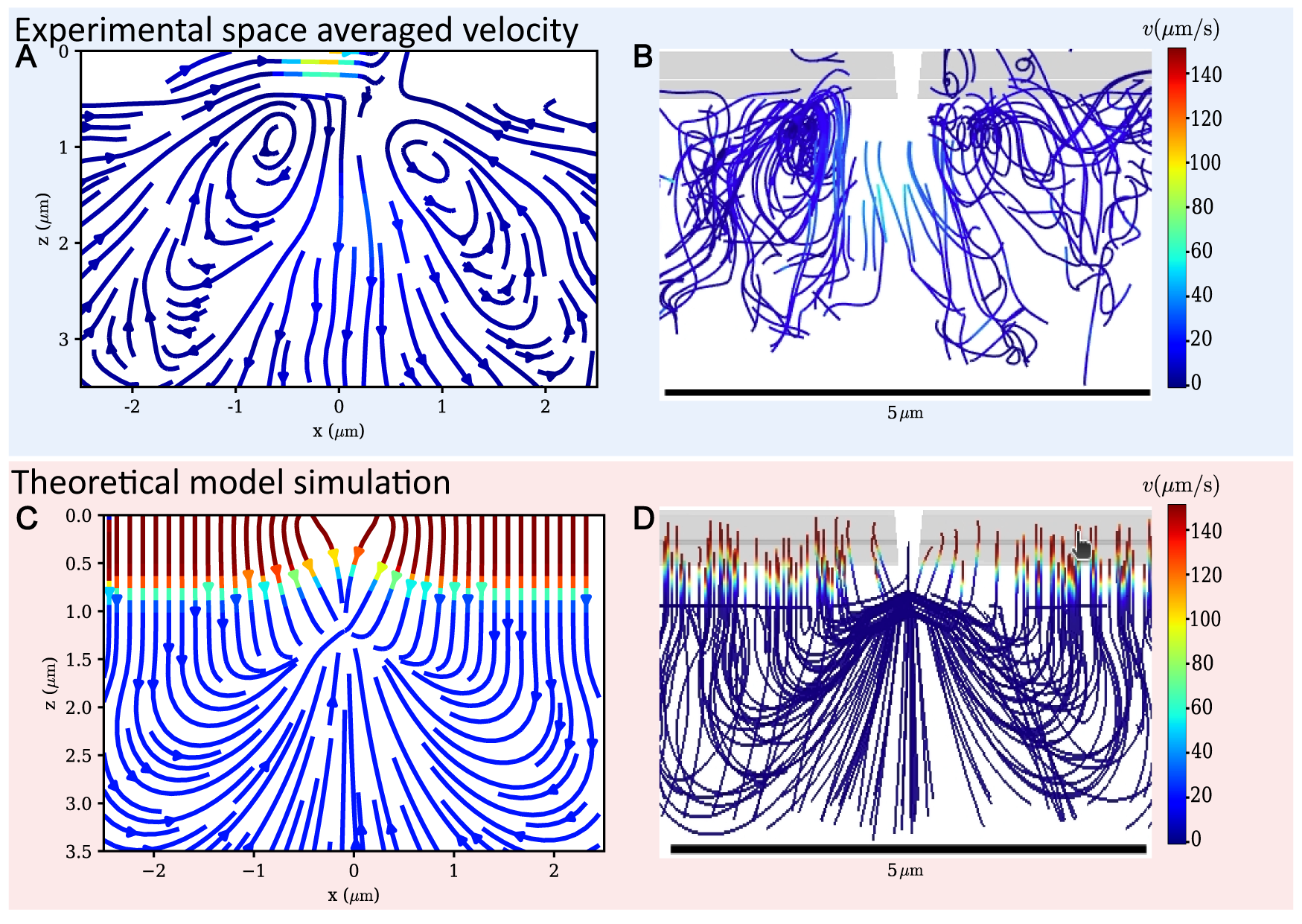}
    \caption{Comparison of 3D flows in nDEP observed by experiment and predicted by theory: A) Streamline calculated from spatially averaged velocities in a cross section taken perpendicular to the electrode corridor. B) Full 3D view of the streamline. C) Theoretical model of nDEP force and electrokinetic flow, projected on the same axis as panel A. D) Full 3D representation of the calculated flow lines in nDEP.}
    \label{fig:6}
\end{figure}
Figure \ref{fig:6} shows the same information as Figure \ref{fig:4}, but for nDEP. In Figures \ref{fig:6}A and \ref{fig:6}B, we can see the vortices that appear in the experimental analysis. The direction of these vortices is clockwise, opposite to the electroosmotic flows. The 3D view in Figure B reveals that the vortices we have observed in experiment appear only along the direction at 45$^o$ with respect to the electrode arms, and that most of the dynamical information in nDEP comes from the center of the quadrupole. In Figures C and D, we can see that neither the DEP force nor the EK flows included in our model can explain the experimentally observed dynamics.

\section{Conclusions}
The data presented in this manuscript represents the first example of combined 3D mapping and computational simulation of the time-dependent flow and velocity field of nanoparticles under positive and nDEP conditions. This work demonstrates the importance of access to a complete experimental dataset for validation of a theoretical model. As discussed higher, 3D NP dynamics close to a quadrupolar electrode generating an AC field revealed rather complicated patterns mainly arising from the combined effect of electrokinetics, pDEP and  particle diffusion.  pDEP attracts particles to the electrode edges, inside the corridor between the electrodes. The observed 3D trajectories of particles getting trapped are reproduced very well by including ion-slip induced EK flows in the theoretical model. The particles going to the trap follow a semi-hyperboloid profile, while other particles that are initially attracted subsequently get ejected, following a circular, vortex-like trajectory due to electro-osmotic flows. In nDEP on the other hand particles are repelled from the gaps between the electrodes (pushing them downwards) and this results in a depletion of particles directly above the gap beween the electrodes. However, particles are attracted to a region about 500 nm deep in the solution, along the edges of the electrodes. The model incorporating DEP and electro-osmotic flow does not fully capture the observed behavior, indicating the importance of other effects in this regime. For example, the inclusion of electro-thermal effects in the theoretical description could be of interest to further investigate small particles under nDEP conditions, as it could lead to the accumulation of particles close to the electrodes. These discrepancies only become clear at the level of detail provided by our joint theory-experiment approach: The prime use of multiscale modeling combined with fast 3D multiplane microscopy imaging enabling direct superresolved visualisation of particle trajectories and subsequent analysis of particle velocities over a 3D volume. In the future, quantitative analysis of 3D particle speeds combined with a more accurate determination of electrokinetic flows using theory will reveal the 3D DEP forcefield, leading to a further understanding of the mechanisms driving these dynamics and offering valuable insights for future applications in nanoparticle manipulation and assembly. The unprecedented level of detail attainable with the presented approach is highly valuable when combined with theory and simulations specifically adapted to modeling at a similar system scale, targeting simulation outputs that can directly be compared with experiment. Building on this approach, we expect to be able to measure the particle polarizability under different conditions and elucidate the details of electrokinetic flow and its interaction with dielectric particles in dense suspensions, where particle-induced EK flows might become relevant, as well as other types of interactions. Moreover, this method is highly promising for the study of sub-100 nm particles, where the assumptions inherent in the Clausius-Mossotti equations might no longer be valid. A first proof of concept was provided by imaging 63 nm particles.  In conclusion, we have provided a comprehensive experiment and modeling framework to study nanoparticle manipulation and flows, adaptable to any electrode geometry and broadly applicable to colloidal systems in general. 

\section{Materials and methods}
\emph{Multiplane widefield microscope}
All experiments have been performed using a multiplane widefield microscope (Figure 1) Briefly, a 488 or 561 nm laser line was used as widefield photoexcitation source by focusing it at the back focal plane of a water-immersion objective lens (N.A.=1.20, 60x, Olympus UplanSApo60XW). The fluorescence emission was collected by the same objective lens and filtered with the suitable optical filter(s) to remove the widefield excitation laser and the emission of the unwanted fluorescent MPs. Afterwards, the emitted light went through a set of lenses in telecentric 4f configuration [Objective - Lens1 – field stop - Lens2 - Tube lens – Camera(s)]. The field stop was used to control the size of the image. Between the tube lens and the cameras, we placed a proprietary prism, which split the entering photon flux into eight different beams with slightly different optical path lengths. The distance between consecutive planes is roughly 580 nm, yielding an axial range of approximately 4 $\mu$m. The different imaging planes were recorded by 2 CMOS cameras (4 imaging planes for each camera, Orca Flash 4.0, Hamamatsu Photonics Inc.). To ensure fast acquisition rates (100 fps) and correct synchronization of the two CMOS cameras, we triggered the acquisition via a National Instruments board (NI, USB-6343) coupled to a home-built software written in Labview. The imaging volume was 400 pixels x 400 pixels x 8 planes (40 × 40 × 4 $\mu$m $^{3}$).\\

\emph{Single Particle Tracking (SPT) analysis}
To track the motion of 200nm diameter PS Fluospheres (Thermofischer, we used a house-written algorithm. The detail of the algorithm used for detecting and tracking the NPs is described elsewhere.23 Briefly, each NP is detected on different planes and then localized in 3D. After calculating the (x-, y-, z-) coordinates of each single emitter in the time-lapse image sequence, they were connected in time to form trajectories. The correspondence assignment was performed by minimizing the sum of the displacements of all NPs via Munkres algorithm. Both localization and tracking algorithms are freely available at https://github.com/BorisLouis/3D-image-processing.\\

\emph{Sample preparation for microscopy}
Since the multiplane widefield microscope (MPM) is an inverted microscope, we used a circular spacer (Grace Bio-Labs SecureSeal™ imaging spacer, diameter 9 mm, thickness 0.12 mm) to create a small sample chamber on the surface of the chip, which holds about 8 $\mu$ L of sample solution, and is closed off with a 0.15 mm cover glass. Commercially available fluorescently labelled carboxylate polystyrene (PS) FluoSpheres™ Fluorescent Microspheres (200 nm) s (Invitrogen) were diluted 10mM MES buffer solution (pH=6), enabling control of surface charges and ionic strength. The procedure to prepare the buffer was as follows: 1.952 g MES free acid was dissolved in 800 ml deionized water. 0.442 g KCl was added to fix the ionic strength at 10 mM. Then the final volume of the solution was adjusted to 1000 ml with deionized water. Then the solution was titrated to pH 5.99 with monovalent strong base or acid as needed. \\

\emph{Device processing}
The polynomial design, known as quadruple, is fabricated using the standard silicon processing techniques. Initially, the 1000 nm silicon oxide was deposited on the P-type 300 mm silicon wafer. This was followed by the deposition of a 2 nm TaN and 50 nm Ruthenium metal layer. The electrode shape was further processed on this metal layer with lithography patterning and etching. Next, a 20 nm silicon nitride layer was deposited, followed by the deposition of a 50 nm silicon oxide. Finally, the etching was further performed to make the Ru metal layer exposed. The center separation of the quadruple is 2 µm. The design generates a steep electric field gradients at the tips and edges, while simultaneously generating a steep electric field that points towards low electric field at the center of the quadruple.

\section{Acknowledgments}
This work was funded by the European Union under the Horizon Europe grant 101130615 (FASTCOMET). This work was also supported by the Flemish Government through long term structural funding Methusalem (CASAS2, Meth/15/04), by the Fonds voor Wetenschappelijk Onderzoek-Vlaanderen (FWO, W002221N), by a bilateral agreement between FWO and MOST (VS00721N), by the internal funds of KU Leuven (C14/22/085), by the Spanish Agencia Estatal de Investigación and FEDER (PID2022-137569NA-C44, PID2022-137569NB-C43, PID2020-117080RB-C51 and PDC2021-121441-C21). B.L. acknowledges FWO for his junior postdoctoral fellowship (12AGZ24N). R.B.-O. thanks the Spanish Agencia Estatal de Investigación for a Ramon y Cajal contract (RYC2021-032773-I). M.I.M. acknowledges financial from the Spanish Ministry of Science and Innovation (MCIN), AEI, and FEDER (UE) through Project No. PID2022-137569NB-C43 and the María de Maeztu Programme for Units of Excellence in R\&D (CEX2023-001316-M).1
The author thanks Geert Van den bosch, Senne Fransen, Wim Van Roy, Kruti Trivedi, Radin Tahvildari and Maarten Rosmeulen, for the fabrication of the electrodes, discussions and support in the framework of the FASTCOMET project.

\section{Abbreviations}
DEP dielectrophoresis; pDEP positive dielectrophoresis; nDEP negative dielectrophoresis; EK electrokinetics; MPM multiplane widefield microscopy; NP nanoparticle
\bibliography{merge}
\end{document}